\documentstyle[aps,eqsecnum,epsfig]{revtex}

\tighten
\begin{document}
\draft
\twocolumn[\columnwidth\textwidth\csname@twocolumnfalse\endcsname

\title{Structure of $^{19}$C from Coulomb dissociation studies}

\author{P. Banerjee and R. Shyam}
\address{Theory Group, Saha Institute of Nuclear Physics,
\\1/AF Bidhan Nagar, Calcutta - 700 064, INDIA}
\date{\today}

\maketitle

\begin{abstract}
We investigate the structure of the neutron rich nucleus $^{19}$C
through studies of its breakup in the Coulomb field of target nuclei.
The breakup amplitude is calculated within an adiabatic treatment
of the projectile excitation, which allows the use of the realistic
wave functions for the relative motion between the fragments in the
ground state of the projectile.  
The angular distribution of the center of mass of $^{19}$C, longitudinal 
momentum distribution of $^{18}$C  
and relative energy spectrum
of the fragments (neutron - $^{18}$C) following the breakup of
$^{19}$C on heavy targets at beam energies 
below 100 MeV/nucleon have been computed using different configurations
for the ground state wave function of $^{19}$C. 
In all the cases, the data seem to favor a
$^{18}$C(0$^+)\otimes$1$s_{1/2}$
configuration for the ground state of $^{19}$C, with the one-neutron
separation energy of 0.53 MeV.   
\end{abstract}
\addvspace{5mm}]
\vfill

The neutron rich nucleus $^{19}$C (the last bound odd-neutron
isotope of carbon) is a strong candidate for having a one-neutron halo
structure \cite{baz95,mar96}, due to its very small one-neutron
separation energy ($S_n$)
(of the order of a few hundred keV). Several measurements reported recently
do seem to provide evidence in favor of such a possibility. The measured 
longitudinal momentum distributions of $^{18}$C fragment emitted in the 
breakup reaction of $^{19}$C on Be and Ta targets at the beam energy
of 88 MeV/nucleon, and on a carbon target at $\sim$ 1 GeV/nucleon,
have widths (full width at half maximum (FWHM)) of 42 $\pm$ 4 MeV/c,
41 $\pm$ 3 and  69 $\pm$ 3 MeV/c respectively \cite{baz95,bau98}.  
The neutron momentum distribution, 
measured in a core-breakup reaction of $^{19}$C at 30 MeV/nucleon on a 
Ta target at GANIL, shows a FWHM of 64 $\pm$ 17 MeV/c \cite{mar96}.
This is about three times smaller than that predicted by the Goldhaber
model~\cite{gold74}, which provides a good description of this quantity
in the case of stable nuclei. 
The measured interaction cross section \cite{oza98} of $^{19}$C on a $^{12}$C 
target at beam energy of $\sim$ 960 MeV/nucleon (1231 $\pm$ 28 mb)
is seen to be enhanced as compared to that of $^{18}$C (1104 $\pm$ 15 mb),
which signals a larger matter radius of $^{19}$C \cite{tos99}. Very
recent measurement of the relative energy spectrum of $^{19}$C in its
Coulomb dissociation at 67 MeV/nucleon on a Pb target \cite{nak99} 
shows a strong peak at very low relative energy, which is supposed to be a
characteristic of the halo structure \cite{han95}.

However, there are still some open issues which make the existence
of a one-neutron halo structure in $^{19}$C somewhat unsettled. The
detailed structure of this nucleus remains uncertain, partly because
of the large uncertainty in its mass \cite{mar96,vie86,gil87,wou87,orr91}.
The latest Nubase evaluation \cite{aud97} gives the value of
$S_n$ = 160 $\pm$ 110 keV.
However, analysis of the data on the interaction cross
section of $^{19}$C \cite{tos99} and relative energy spectrum of
fragments observed in the Coulomb breakup of $^{19}$C \cite{nak99} seem to 
favor a $S_n$ $\simeq$ 0.5 MeV (with a particular configuration for the
$^{19}$C ground state). Moreover, in a recent measurement of the 
neutron angular distribution in the elastic breakup of $^{19}$C 
(in which target nucleus remains in the ground state),
a broad FWHM  of 120 $\pm$ 18 MeV/c has been reported \cite{lie98}. This
value is about two times larger than the results reported in
Ref.\cite{mar96}, wherein a narrow width was observed.

In this paper, we analyze the available data on the breakup
of $^{19}$C on heavy target nuclei within the framework of a theory
of the breakup reactions which allows the use of realistic wave functions
to describe the projectile ground state. By comparing
our calculations (performed with various configurations for the 
$^{19}$C ground state) with the data, we hope to put 
a constraint on the ground state structure of this nucleus. 
This is expected to clarify the issue regarding the existence of
a halo structure in $^{19}$C since the spin-parity and the
configuration of the valence neutron play an important role in the
formation of the halo. We shall consider here only the Coulomb breakup
process, which dominates the breakup of the loosely bound projectiles
on heavy target nuclei~\cite{prab93} in kinematical regime below the
grazing angle.

The theory of the Coulomb breakup (CB) of the projectile used by us
has been described extensively in \cite{ban98,tos98}. The triple
differential cross section for the reaction, in which a projectile $a$ 
breaks up into a charged core $c$ and a neutral valence particle $n$ on 
target $t$, is given by
\begin{eqnarray}
{d^3\sigma \over dE_c d\Omega _cd\Omega _n} & = &
{2\pi\over \hbar v_a}\left\{\sum_{l
\mu}\frac{1}{(2l + 1)}\vert \beta^{\rm AD}_{l\mu}\vert^2 \right\}\nonumber\\
                  & & \rho(E_c, \Omega _c,\Omega _n)~.
\end{eqnarray}
Here $v_a$ is the $a$--$t$ relative velocity in the entrance channel and
$\rho (E_c,\Omega _c,\Omega _n)$ the phase space factor appropriate 
to the three-body final state.  The amplitude $\beta^{\rm AD}_{l\mu}$
is given by
\begin{eqnarray}
\beta^{\rm AD}_{l\mu}
& = &\langle \vec{q}_n\vert V_{cn}\vert \Phi _{a}^{l\mu}\rangle
\langle \chi ^{(-)}(\vec{k}_c);\alpha\vec{k}_n\vert \chi
^{(+)}(\vec{k}_a) \rangle~,
\end{eqnarray}
where $\vec{q}_n = \vec{k}_n -\gamma \vec{k}_a$, with
$\gamma = m_n/(m_c+m_n)$.  
$\vec{k}_n$ and $\vec{k}_a$ are the asymptotic momenta of neutron
and projectile
and $m_n$ and $m_c$ are the masses of neutron and the core. 
The first term in Eq. (2) contains the structure information about
the projectile through the
ground state wave function $\Phi _{a}^{l\mu}(\vec{r})$, and it is
known as the vertex function \cite{ban98}, while the second
term is associated only with the dynamics of the 
reaction, which can be expressed in terms of the
bremsstrahlung integral \cite{nor54}. For the 
explanation of other quantities in the above equation, we
refer to \cite{ban98}.

The CB theory is fully quantum mechanical and is also non-perturbative.
The method retains finite-range effects associated with the interaction
between the breakup fragments and includes the initial and final state
Coulomb interactions to all orders. It allows the use of wave functions
of any relative orbital angular momentum for the motion between
$c$ and $n$ in the ground state of $a$. It should, however, be mentioned
that to obtain Eq. (2) \cite{tos98}, it has been assumed that the dominant
projectile breakup configurations excited are in the low-energy continuum
(the adiabatic approximation). Furthermore, this theory is not applicable 
to those cases where the valence particle is charged. 

Expression for the Coulomb breakup amplitude in the factorised form,
which also uses the bremsstrahlung integral, has been obtained
previously \cite{shyam92} within the distorted wave Born approximation
(DWBA), by making the approximation of replacing the vector describing 
the separation of $c+n$ c.m. with respect to the target by that of
$c$ with respect to target in the projectile wave function. The important
difference between Eq. (2) and the corresponding DWBA expression so
obtained is that in the latter the vertex function is evaluated
at momentum $\vec{k}_n$ instead of $\vec{q}_n$. It has been shown \cite{tos98}
that DWBA with this approximation underestimates the ($d,pn)$ breakup
cross sections at beam energies $\geq$ 140 MeV.

In the calculations of the vertex function, we have considered 
the following configurations for the valence neutron in the $^{19}$C
ground state: (a) a 1$s_ {1/2}$ state bound to a 0$^+$ $^{18}$C core
by 0.24 MeV, (b) a 1$s_{1/2}$ state bound to a 2$^+$ $^{18}$C core
by 1.86 MeV, (c) a 0$d_{5/2}$ state bound to a 0$^+$ $^{18}$C core
by 0.24 MeV, and (d) a 1$s_{1/2}$ state bound to a 0$^+$ $^{18}$C core
with 0.53 MeV. The authors of Ref.~\cite{nak99} use option (d) with
a spectroscopic factor (SF) of 0.67. 
The binding potentials in all cases are taken to be of 
Woods-Saxon type having central and spin-orbit terms with the
radius and diffuseness parameters being 1.236 fm and 0.62 fm
respectively. The strength of the spin-orbit term was held fixed to
7 MeV \cite{nak99}. The depths of the central potentials were
searched so as to reproduce the respective binding energies.
The rms sizes of $^{19}$C with options (a)-(d) were found to be 
3.50 fm, 3.03 fm, 3.00 fm and 3.23 fm respectively, while the 
corresponding rms sizes of the valence neutron were 9.26 fm, 4.93 fm,
4.43 fm and 7.07 fm respectively. The rms size used for the $^{18}$C 
core is 2.9 fm \cite{lia90}. These different wave functions of
$^{19}$C give rise to different vertex functions and consequently,
different Coulomb breakup cross sections. 

In Fig. 1, we present a comparison of our CB model calculations (performed
with configurations (a)-(d) as mentioned above) and the
experimental data for the angular distribution of the center of mass (c.m.) of 
the $n$ + $^{18}$C system in the breakup of $^{19}$C on a Pb target at the beam 
energy of 67 MeV/nucleon \cite{nak99}. The integration over the relative energy
is performed in the range of 0 - 0.5 MeV (the same as is done in
Ref.~\cite{nak99}). We see that out of the four cases, the calculation
done with option (d) (with a SF = 1.0) (solid line) is in the best agreement 
with the data. Still, the quality of the fit to the data is not as good as 
that seen in Ref.~\cite{nak99}. This is due to the fact that in
Ref.~\cite{nak99} the calculated cross sections have been folded with
the experimental angular resolution. After performing a similar folding,
the agreement between the CB model results (shown by the dash-dotted line)
and the data is of the similar nature as that seen in Ref.~\cite{nak99}.
It may, however, be noted that the semiclassical Coulomb excitation (SCE) 
calculations reported by these authors \cite{bar88}
use option (d) with a SF of 0.67, instead of 1.0 as used by us.
With our choice of SF, the SCE theory overpredicts the experimental
cross sections. 
\begin{figure}[tbh]
\begin{center}
\epsfxsize=8.3cm
\epsfbox{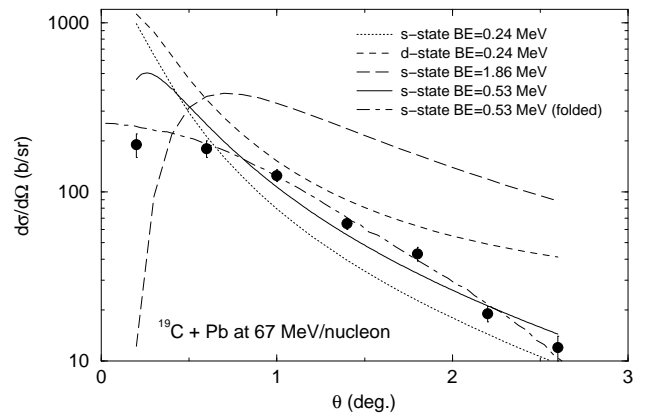}
\end{center}
\caption[C1] {Calculated angular distribution of the centre of mass of
$^{19}$C in its Coulomb breakup on a Pb target at 67 MeV/nucleon. The dotted, 
short-dashed and long-dashed curves have been multiplied by 0.2, 100 and 100
respectively. The dash-dotted line is the result of the calculation performed 
with option (d) folded with experimental angular resolution. The experimental
data in taken from \protect\cite{nak99}.} 
\label{fig:figa}
\end{figure}
\noindent
{}In Fig. 2, we compare the results of our calculations
with the experimental data for the relative energy ($E_{rel}$) 
spectrum of the fragments for the same reaction as in Fig. 1. 
As in Ref.~\cite{nak99}, the angular integrations for the $^{19}$C c.m.
are done up to the grazing angle of 2.5$^{\circ}$. We note that in this case 
too, the best agreement with the data (near the peak position) is obtained
with the configuration (d) (with SF = 1.0) for the $^{19}$C ground state. 
Here again, the SCE calculations done with the same configuration
and SF overestimate these cross sections. A SF of 0.67 is required  
to make the SCE results (shown by the dash-dotted line) 
agree with the data in the peak region. One reason for this
difference in the CB and SCE results could be the fact that in 
latter the energy distribution of the dipole excitation ($B(E1))$
has been obtained with a plane-wave to describe the relative motion 
between the fragements in the final state. Consideration
of the nuclear interaction for this may lead to a
reduction in the cross section as is observed for the case of
$^8$B~\cite{nunes96}. It may be noted that in the CB model the final
state consists only of the product of the wave functions describing 
the $c$+target and $n$+target relative motions;
the wave function for the relative motion between the
fragments does not enter here. We use a  
plane wave to describe the neutron-target relative motion 
which is valid for the case of pure Coulomb breakup of projectiles with
a chargeless fragment in the outgoing channel.

We would like to emphasize that, the SCE and CB theories use entirely
different mechanisms to describe the Coulomb breakup process. The CB
theory uses the post-form scattering amplitude which includes breakup
contributions from entire continuum corresponding to all Coulomb multipoles
and relative orbital momenta between the valence and core fragments.
In contrast to this, the SCE calculations include contributions from one
(as is the case in Ref.~\cite{nak99}) or at the most two multipolarities.
\begin{figure}[tbh]
\begin{center}
\epsfxsize=8.3cm
\epsfbox{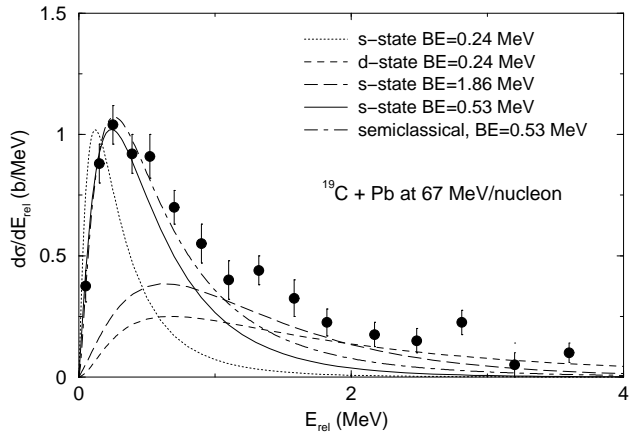}
\end{center}
\caption[C1] {Calculated relative energy spectra in the Coulomb dissociation
of $^{19}$C on a Pb target at 67 MeV/nucleon. The dotted, short dashed and
long dashed curves are results of multiplication by 0.16, 10 and 10
respectively of the actual calculations. The dash-dotted line is the result
of the semiclassical calculation performed with option (d) (see text). 
For the semiclassical calculation, we use SF = 0.67.
The experimental data have been taken from \protect\cite{nak99}.}
\label{fig:figb}
\end{figure}
\noindent
{}Furthermore, often the continuum is curtailed to some maximum value.
Therefore, the differences seen in the SCE and CB models 
should not be surprising. 

Of course, both the CB and SCE theories underestimate the relative
energy spectrum for larger values of $E_{rel}$. Although, the SCE 
results are comparatively better in this regard. Improper consideration
of the nuclear breakup effects could be one of the reasons for this
disagreement. Indeed, Dasso {\em et al.}~\cite{dasso99} have shown that 
nuclear breakup cross sections dominate the relative energy spectrum of
fragments for $E_{rel}$ $>$ 0.6 MeV in case of the breakup of
$^{11}$Be (also a one-neutron halo nucleus) on a Pb
target at a similar beam energy (72 MeV/nucleon). The authors of Ref.
\cite{nak99} do correct their data for the nuclear breakup effects 
by scaling the corresponding cross sections for the breakup of
$^{19}$C on a carbon target. However, this is unlikely to be 
accurate as the scaling procedure may not be valid due to the long
range of the nuclear interaction in the halo nuclei~\cite{dasso98}. 

The CB and SCE calculations (with configuration (d) and SF = 1.0)
predict a total Coulomb breakup cross section of 0.78 b and
1.53 b respectively. The latter is even larger than the experimental
value of the total breakup cross section of 1.34 $\pm$ 0.12 b \cite{nak99}. 
This again underlines the
necessity of using a SF of 0.67 in the SCE calculations.  As far as
the CB model results are concerned, one should keep in mind that
there are non-negligible nuclear breakup cross sections, which must be
considered together with it for any comparison with the experimental   
total breakup cross sections. In a quantum mechanical theory, however,
one has to add coherently the amplitudes of the Coulomb and nuclear
breakup processes. Therefore, it is not correct to simply add the
cross section of the two processes calculated separately 
for comparison with the data on total breakup. Unfortunately, the
extension of the CB theory to include the nuclear breakup
effects~\cite{ron98} is non-trivial and has not been attempted so far.
\begin{figure}[tbh]
\begin{center}
\epsfxsize=8.3cm
\epsfbox{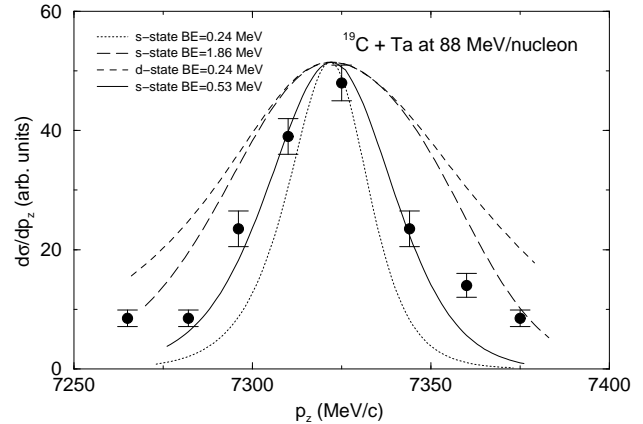}
\end{center}
\caption[C1]{Calculated parallel momentum distributions of $^{18}$C
from Coulomb breakup of $^{19}$C on Ta at 88 MeV/nucleon. The calculations
have been shifted to the data to compare the widths and the peaks
normalized to the peak of the data. The experimental data are taken
from \protect\cite{baz95}.} 
\label{fig:figc}
\end{figure}
\noindent
{}The parallel momentum distribution (PMD) of the heavy charged fragment
(which provides an almost unambiguous information on the existence of a
halo structure in the projectile \cite{baz95,ban98,orr95,kel95,ban95,bar92})
is expected to have a large contribution from the Coulomb breakup process
in the peak region\cite{baz95,ban98}. 
In Fig. 3, we show the comparison of our calculations with data~\cite{baz95}
for the PMD of the $^{18}$C fragment emitted in the 
breakup of $^{19}$C on a Ta target at 88 MeV/nucleon beam energy. 
The results obtained with options (a), (b) and (c) are similar to  
those shown in Ref.\cite{ban99}. The MSU data do not have the absolute
magnitudes of the PMDs. Therefore, we have normalized the peaks of the
calculated PMDs to that of the data (this also involves the shifting of
the position of the maxima of calculations to those of the data), so that  
the widths of the calculated and the measured distributions can be compared.

The FWHM of the PMDs calculated with the options
(a), (b), (c) and (d) are 27, 71, 83 and 41 MeV/c respectively, while that
of the experimental one is 41 $\pm$ 3 MeV/c. Thus the option (d)
is favoured by this data as well. It should be noted that 
in a study of the breakup of $^{19}$C on a Be target \cite{rid97}
within a core-plus-neutron coupling model with a deformed
Woods-Saxon potential for the neutron-core interaction, an agreement
with the FWHM of the experimental PMD has been obtained with the
configuration $^{18}$C(0$^+)\otimes$1$s_{1/2}$ (with $S_n$ = 0.5 MeV), 
for the ground state of $^{19}$C. 
Although this result is for a lighter target  
(Be), for which Coulomb breakup is not the dominant reaction 
mechanism, yet its comparison with that of ours may
not be out of the place as the PMD is least affected by the
reaction mechanism \cite{kel95,ban96}. 

In conclusion, we have studied Coulomb breakup of the neutron rich exotic
nucleus $^{19}$C and compared our calculations with the existing experimental 
data in order to probe its ground state structure. The breakup amplitude is
calculated within an approximate quantum mechanical theoretical model, 
which assumes that the important excitations of the projectile are to
the low-energy continuum so that they can be treated adiabatically. 
The method permits a finite-range treatment of the projectile vertex and
includes initial and final state Coulomb interactions to all orders.
 
We find that the calculations performed with the configuration
$^{18}$C(0$^+)\otimes$1$s_{1/2}$ for the ground state of $^{19}$C
(with a one-neutron separation energy of 0.53 MeV) agree
with the data on the angular distribution of the $^{19}$C c.m.,
parallel momentum distribution of the breakup 
fragment $^{18}$C and the relative energy spectrum of $^{19}$C,
better than those done with other configurations. Therefore, 
these data seem to support this configuration as the dominant
component in the ground state wave function of
$^{19}$C. This also gives credence to the existence of a  
neutron halo structure in this nucleus. 

Calculations done within a semiclassical Coulomb
excitation theory (with the same configuration) 
overestimate the magnitudes of the angular distribution of the
$^{19}$C c.m. and the relative energy spectrum of the breakup fragments
in the peak region; a spectroscopic factor
of 0.67 is required to explain the data within this model. 
The use of a plane wave to describe the relative
motion between the fragments in the final state in obtaining  
the energy distribution of the dipole excitation used in these 
calculations may be one of the reasons for this overestimation.
It would be worthwhile to redo these calculations by considering the
nuclear distortion effects in the final channel. 
Proper consideration of the nuclear breakup effects in both (semiclassical
Coulomb excitation as well as Coulomb breakup) models is also necessary
to explain the relative energy spectrum at higher relative energies.

The authors would like to thank Takashi Nakamura for several helpful
correspondences and for performing the folding of the calculated cross
sections shown in Fig. 1.


\begin{references}
\bibitem{baz95}
D. Bazin {\em et al.}, Phys. Rev. Lett. {\bf 74}, 3569 (1995); 
Phys. Rev. {\bf C57}, 2156 (1998).

\bibitem{mar96}
F. M. Marqu\'es {\em et al.}, Phys. Lett. {\bf B381},
407 (1996).

\bibitem{bau98}
T. Baumann {\em et al.}, Phys. Lett. {\bf B439}, 256 (1998).

\bibitem{gold74}
A.S. Goldhaber, Phys. Lett. B {\bf 53} (1974) 306;
D.J. Morrissey, Phys. Rev.C {\bf 39} (1989) 460.

\bibitem{oza98}
A. Ozawa {\em et al.}, RIKEN preprint RIKEN-AF-NP-294,
August 1998, submitted to Physical Review Letters.

\bibitem{tos99}
J. A. Tostevin and J. S. Al-Khalili, Phys. Rev. {\bf C59}, R5 (1999).
\bibitem{nak99} T. Nakamura {\em et al.}, Phys. Rev. Lett. {\bf 83}, 1112 
(1999) and private communication.
\bibitem{han95} P.G. Hansen, A.S. Jensen and B. Jonson, Ann. Rev. Nucl. Part.
Sci. {\bf 45}, 2 (1995).
\bibitem{vie86} D. J. Vieira {\em et al.}, Phys. Rev. Lett. {\bf 57},
 3253 (1986).
\bibitem{gil87} A. Gillibert {\em et al.}, Phys. Lett. {\bf B192}, 39 (1987).
\bibitem{wou87} J. M. Wouters {\em et al.}, Z. Phys. {\bf A331}, 229 (1988).
\bibitem{orr91} N. A. Orr {\em et al.}, Phys. Lett. {\bf B258}, 29 (1991).
\bibitem{aud97} G. Audi {\em et al.}, Nucl. Phys. A {\bf 624}, 1 (1997).
\bibitem{lie98} E. Liegard, N. A. Orr {\em et al.}, LPC-Caen report LPCC
98-03, to be published; E. Liegard, Th\`ese, Universit\'e de Caen, France
(1998), unpublished.
\bibitem{prab93} P. Banerjee and R. Shyam, Nucl. Phys. A {\bf 561} (1993) 112.
\bibitem{ban98} P. Banerjee, I. J. Thompson and J. A. Tostevin, Phys. Rev.
 {\bf C58}, 1042 (1998).
\bibitem{shyam92} R. Shyam, P. Banerjee and G. Baur, Nucl. Phys. A
{\bf 540} 341, (1992).
\bibitem{tos98} J. A. Tostevin {\em et al.}, Phys. Lett. {\bf B424},
 219 (1998); J. A. Tostevin, S. Rugmai and R. C. Johnson, Phys. Rev.
 {\bf C57}, 3225 (1998).
\bibitem{nor54} A. Nordsieck, Phys. Rev. {\bf 93}, 785 (1954).
\bibitem{lia90} E. Liatard {\em et al.}, Europhys. Lett. {\bf 13}, 401 (1990).
\bibitem{bar88} C. A. Bertulani and G. Baur, Phys. Rep. {\bf 163}, 299 (1988).
\bibitem{nunes96} F.M. Nunes, R. Crespo and I.J. Thompson, Nucl. Phys. A
{\bf 615}, 69 (1997).
\bibitem{dasso99} C. H. Dasso, S. M. Lenzi and A. Vitturi, Phys. Rev. {\bf C59},
539 (1999).
\bibitem{dasso98} C. H. Dasso, S. M. Lenzi and A. Vitturi, Nucl. Phys. 
{\bf A639}, 635 (1998).
\bibitem{ron98} R. C. Johnson, J. Phys. {\bf G24}, 1583 (1998).
\bibitem{orr95} N. A. Orr {\em et al.}, Phys. Rev. C {\bf 51}, 3116 (1995).
\bibitem{kel95} J. H. Kelley {\em et al.}, Phys. Rev. Lett. {\bf 74}, 30 (1995).
\bibitem{ban95} P. Banerjee and R. Shyam, Phys. Lett. {\bf B349}, 421 (1995).
\bibitem{bar92} C. A. Bertulani and K. W. McVoy, Phys. Rev. {\bf C46},
 2638 (1992).
\bibitem{ban99} P. Banerjee, J. A. Tostevin, and I. J. Thompson, J. Phys.
 {\bf G25}, 851 (1999).
\bibitem{rid97} D. Ridikas, M. H. Smedberg, J. S. Vaagen, and M. V. Zhukov, 
Europhys. Lett. {\bf 37}, 385 (1997); Nucl. Phys. {\bf A628}, 363 (1998). 
\bibitem{ban96} P. Banerjee and R. Shyam, J. Phys. {\bf G22}, L79 (1996).
\bibitem{ann93} R. Anne {\em et al.}, Phys. Lett. {\bf B304}, 55 (1993).
\end{references}
\end{document}